\documentclass[natbib]{svjour3}             
\smartqed  
\usepackage{graphicx}
\usepackage{natbib}

\begin{document}

\title{How Gaussian competition leads to lumpy or uniform species distributions}



\author{Simone Pigolotti \and Crist\'obal L\'opez \and Emilio
Hern\'andez-Garc\'\i a \and Ken H. Andersen }


\institute{S. Pigolotti \at The Niels Bohr International Academy, the
  Niels Bohr Institute, Blegdamsvej 17 DK-2100, Copenhagen Denmark.
  Tel.: +45-35325238, Fax: +123-45-678910, \email{pigo@nbi.dk} \and C.
  L{\'o}pez and E. Hern{\'a}ndez-Garc{\'\i}a, \at IFISC, Instituto de
  F{\'\i}sica Interdisciplinar y Sistemas Complejos (CSIC-Univ. de les
  Illes Balears), E-07122 Palma de Mallorca, Spain \and K. H. Andersen
  \at National Institute of Aquatic Resources, Technical University of
  Denmark, Charlottenlund slot, DK-2920 Charlottenlund, Denmark.}


\maketitle

\begin{abstract}

  A central model in theoretical ecology considers the competition of
  a range of species for a broad spectrum of resources. Recent studies
  have shown that essentially two different outcomes are possible.
  Either the species surviving competition are more or less uniformly
  distributed over the resource spectrum, or their distribution is
  'lumped' (or 'clumped'), consisting of clusters of species with
  similar resource use that are separated by gaps in resource space.
  Which of these outcomes will occur crucially depends on the
  competition kernel, which reflects the shape of the resource
  utilization pattern of the competing species. Most models considered
  in the literature assume a Gaussian competition kernel. This is
  unfortunate, since predictions based on such a Gaussian assumption
  are not robust. In fact, Gaussian kernels are a border case
  scenario, and slight deviations from this function can lead to
  either uniform or lumped species distributions. Here we illustrate
  the non-robustness of the Gaussian assumption by simulating
  different implementations of the standard competition model with
  constant carrying capacity. In this scenario, lumped species
  distributions can come about by secondary ecological or evolutionary
  mechanisms or by details of the numerical implementation of the
  model. We analyze the origin of this sensitivity and discuss it in
  the context of recent applications of the model.

\end{abstract}

\section*{Introduction}
\label{intro}

A central model behind the theoretical description of competition
among dissimilar species was introduced by \citet{MacArthur1967}. In
the model, species are characterized by their niche position $x_i$,
which measures a trait being relevant for the exploitation of a
distributed resource.  As examples, the niche value $x_i$ may
represent body size of predators, where the distributed resource are
preys with their size distribution, or $x_i$ could be beak sizes of
birds, in which case the resource could be seeds of different
sizes. Mathematically this leads to a Lotka-Volterra type of
competition equation, where the competition coefficients are function
of the distance between species on the niche axis $x$. This
competition kernel is usually taken to be a Gaussian function of the
niche difference (also called normal curve). The implication of this
choice is the central topic of this paper.

The model was originally proposed as part of the hypothesis of
limiting similarity, namely that competing species can coexist only if
they are sufficiently different from each other
\citep{MacArthur1967,Abrams1983}. A mathematical analysis of the model
revealed that arbitrarily similar species could in fact coexist in
some cases. However adding further effects to the model, like noise
(\citet{May1972}, but see \citet{Turelli1978}) or extinction
thresholds \citep{Pigolotti2007}, impose a limit to the similarity
between species. This sensitivity to second order effects has led to
the conclusion that the model, in its original form, is structurally
unstable when used to predict limits of similarity
\citep{Meszena2006}. The competition model has also been applied to
describe coevolving species \citep{MacArthur1967,Case1981} and used in
some formulations of the theory of island biogeography
\citep{Roughgarden1979}. More recently the same type of model has been
simulated numerically and used as a basis for dynamical models of
sympatric speciation \citep{Doebeli2000}, food web assembly and
evolution \citep{Loeuille2005,Johansson2006,Lewis2007}, elucidating
the relation between competition and predator-prey interactions
\citep{Chesson2008}, and for explaining lumped size distributions of
species \citep{Scheffer2006}. For a more extensive review of the
biological applications and the generalization of the model see
\citep{szabo}. The competition model has been fundamental for
the development of basic principles in theoretical ecology and it is
relevant to achieve a full understanding of its technical
aspects.

In almost all applications of the model the chosen competition kernel
is Gaussian. This choice facilitates mathematical analysis, and was
justified because the exact shape of the kernel was thought to have no
influence on the fundamental results of the model. However, recent
work has shown that the equilibrium solution can be one of two
fundamentally different types, depending on the form of the
competition kernel \citep{Pigolotti2007}. One class of competition
kernels preserves all species initially introduced in the system, with
adjustments only in their relative abundance. The final equilibrium is
a state with species closely spaced and with roughly similar
abundances (if the carrying capacity is also uniform). Another class
of competition kernels leads to the species being lumped in dense
groups (in some cases groups are formed by single species), separated
by empty regions on the niche axis. Subsequent invasion of new species
in these `exclusion zones' is not possible due to competitive
exclusion.  The condition for uniform distribution of species is to
have a {\sl positive definite} competition kernel (see definition
below). This criterion is automatically fulfilled when the kernel is
constructed from the overlap of the species utilization of the
resource \citep{Roughgarden1979}. If the kernel is not positive
definite, a lumpy species distribution with exclusion zones
emerges. The concern about this discovery is that, even though the
Gaussian kernel is ecologically sound, it is exactly marginal between
the two regimes. This indicates that numerical inaccuracies and/or
secondary ecological effects may violate the positive definiteness of
the competition kernel and cause a transition from a uniform to a
lumpy species distribution.

The objective of this paper is to raise awareness in the theoretical
ecology community of the potential pitfalls and subtleties associated
with the use of Gaussian competition kernels or other marginal
choices. To illustrate this, the consequences of the marginal nature
of the Gaussian kernel in the competition model are explored, in the
idealized case of a uniform carrying capacity.  First, the sensitivity
to ecologically relevant effects that may lead to lumpy distributions
are examined. Then we examine the sensitivity to the details of
numerical implementation. In the last section, we discuss the
relevance of our results for the applications of the model.

\section*{Model}

The competition model considers $n$ interacting populations, each
utilizing a common distributed resource $x$ according to a utilization
function $u_i(x)$, $i=1,...,n$. The dynamics of the abundance of
species $i$, $N_i$, is described by a Lotka-Volterra set of
competition equations:
\begin{equation}
  \label{eq:LV}
  \dot{N_i} = N_i \left(1 - \frac{1}{K} \sum_{j =1}^n G_{ij} N_j\right), \ \ i=1,...,n,
\end{equation}
where the growth rate (considered to be the same for all species) is
set to one for simplicity, and the carrying capacity $K$ is
uniform. Competition in (\ref{eq:LV}) is described by competition
coefficients $G_{ij}$ which are constructed from the overlap of
utilization functions of competing species
\citep{MacArthur1967,Roughgarden1979}:
\begin{equation}
  \label{eq:interaction}
  G_{ij} = \frac{ \int u_i(x) u_j(x)\, dx }{ \int u_i^2(x)\, dx }.
\end{equation}
A justification of (\ref{eq:interaction}) rests upon considering the
probability that consumer $i$ meets consumer $j$
\citep{Levins1968,Roughgarden1979}.

\begin{figure*}
\includegraphics[width=\textwidth]{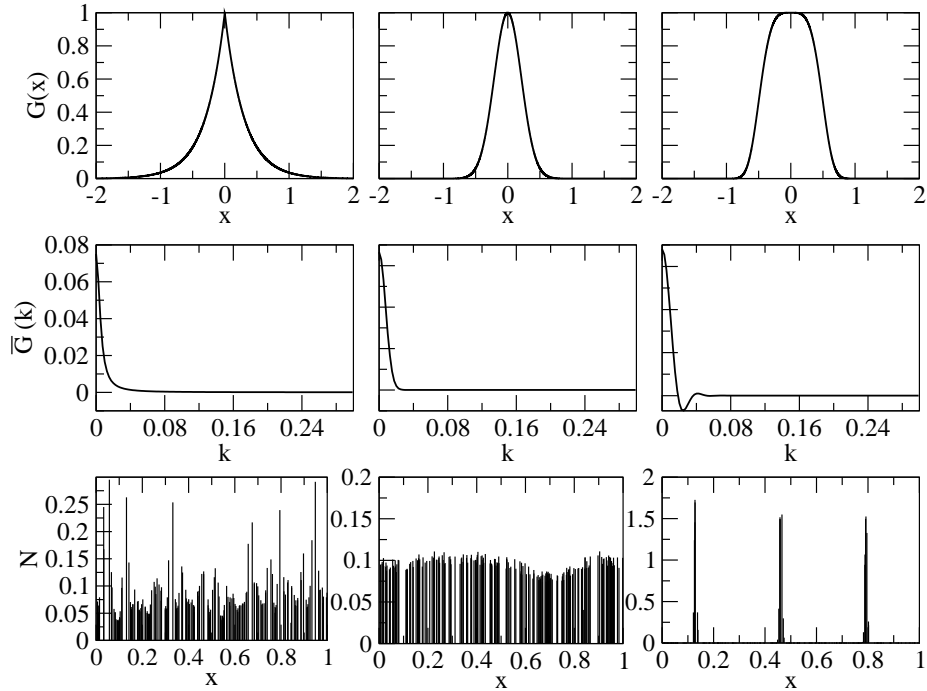}
\caption{Three interaction kernels (top), their Fourier transform
  (middle row), and species distributions arising from simulation of
  the model after 1000 generations (bottom). Column are (Left)
  Exponential competition ($p=1$); (Center) Gaussian competition ($p
  =2$), and (Right) quartic competition ($p=4$). Simulations are
  initiated with $200$ species randomly distributed, $K=10$, and
  $\sigma=0.3$.}
\label{fig:patterns}
\end{figure*}

Often, utilization functions are ignored, and the competition
coefficients are postulated directly. It is usually assumed that
species $i$ has an optimal exploitation of the resource at a value
$x=x_i$, and the competition coefficients are taken to depend on the
difference between the optimal resource values of two competing
species, $y = |x_i-x_j|$, such that we can introduce the so-called
competition kernel, $G_{ij}=G(y)$. Here we use a family of competition
functions described by a parameter $p$:
\begin{equation}
  \label{eq:alpha}
  G_{ij} = G(y)= e^{-|(x_i-x_j)/\sigma|^p},
\end{equation}
which contains the Gaussian kernel when $p=2$, or the exponential one
when $p=1$. The width of the kernel $\sigma$ gives the range of
competition on the niche axis. Incidentally the Gaussian kernel is
obtained from Eq.~(\ref{eq:interaction}) when the utilization
functions are also Gaussian and of the form $u_i=\exp(-((x-x_i)/s)^2)$
with $s^2=\sigma^2/2$. When $p < 2$ the kernels are more peaked around
$y\approx 0$ and for $p>2$ they become more box-like (see
Fig.~\ref{fig:patterns}).

Note that when competition coefficients are constructed by the formula
(\ref{eq:interaction}), i.e.~from the overlap of two utilization
functions, they are always \emph{positive definite}, meaning that
$\sum_{ij} a_i G_{ij} a_j \ge 0$ for any set of numbers $a_i$
\citep{Roughgarden1979} or, equivalently, that the Fourier transform
of $G(y)$, defined as $\tilde{G}(k)=\int_{-\infty}^{+\infty}dx\ 
G(x)\exp(ikx)$, does not take negative values. This property holds for
the family of kernels (\ref{eq:alpha}) for $p \le 2$, but not for
$p>2$ (Fig.~\ref{fig:patterns}). The Gaussian kernel is therefore
marginal in the sense that, corresponding to the limit case $p=2$,
even a very small perturbation may violate its positive definite
character, generally believed to be an ecological requirement arising
from expression (\ref{eq:interaction}).

An intuitive explanation for the appearance of the exclusion zones for
$p>2$ is the following. Interaction kernels with large $p$ have a
box-like shape. In these cases species compete very strongly with
other species, roughly within a distance $\pm \sigma$ from their own
niche value. Species with a niche $x$ in that range will therefore not
be able to invade the resident species, leading to the exclusion zones
between them.  When $p$ is decreased, the resident species compete
less and less with neighbouring species, until the exclusion zones
disappear, leading to the possibility of continuous coexistence.

Understanding the fact that the transition occurs at $p=2$, and also
the coexistence of more than one species in each cluster, requires a
mathematical stability analysis of the model. Consider the uniform
solution, in which many species having the same abundance are densely
packed in niche space. Now perturb each population by a small quantity
$\Delta N_i$, which can be either positive or negative. If the
competition kernel is not positive defined, there are sets of
perturbations such that $\sum \Delta N_j G_{ij} \Delta N_i$ is less
than zero. One can show that such perturbations are amplified by the
dynamics \citep{Pigolotti2007}, making the uniform solution
unstable. The system will then evolve to a clustered state, where the
distance between clusters is proportional to the interaction range
$\sigma$.

We mention here that a possible generalization is to consider
multi-dimensional niche spaces. This possibility would complicate the
mathematical notation but does not introduce qualitative changes.
Stability in a multi-dimensional niche space still depends on the
positive definiteness of the competition kernel.  In particular, a
multi-dimensional Gaussian competition kernel is still marginal in the
sense described above.

We simulated the model (\ref{eq:LV}) with competition kernel
(\ref{eq:alpha}) for 1000 generations and $200$ species initially at
random niche positions. The width of the kernel is $\sigma = 0.3$ and
the carrying capacity is $K=10$.  The niche range is taken to be $x
\in [0,1]$. The standard mathematical way to avoid effects due to the
borders of the niche space is to adopt periodic boundary conditions
(e.g.~\citet{Scheffer2006}). These are introduced for mathematical
convenience and aim at modeling species far from endpoints in a large
niche space. Adopting periodic boundary conditions means that niche
space is circular, so that when the interaction kernels extends beyond
the left edge at $x=0$, it enters back into the right side at $x=1$
and vice versa. Periodic boundaries therefore mimic an infinite system
by considering the niche segment $[0,1]$ as embedded in an array of
repeated copies of itself. Mathematically, this is properly
implemented by making a 'kernel wrap', i.e.~substitute $G(y)$ in
(\ref{eq:alpha}) with $G_p(y)\equiv \sum_n G(y-n)$, where the sum runs
from $n=0,\pm 1, \pm 2, ...\pm\infty$. We call this implementation
``fully periodic boundary conditions'', to distinguish it from another
possibility considered below. Under fully periodic boundary
conditions, the stability properties of the uniform solution are the
same as in the infinite system.

\section*{Results}

Simulations using the competition kernel (\ref{eq:alpha}) with $p= 1$
(exponential), $2$ (Gaussian) and $4$ (box-like) illustrate the
uniform species distributions for $p=1$ and $p=2$, and the lumped
species clusters for $p=4$ (Fig.~\ref{fig:patterns}). The
configurations in Fig.~\ref{fig:patterns} are still transient states
and, at longer times, configurations with $p\le 2$ become more
uniform, whereas the periodically spaced clusters of species for
$p>2$ become thinner until they contain only a single species. 
In any case, from the initial stages until the final equilibrium, the
main difference between the dynamics for the two classes of
competition kernel is unchanged: for $p\le 2$ all initial species are
preserved, leading to dense and evenly distributed configurations,
whereas `exclusion zones' develop for $p>2$ leading to lumped species
distributions.

\subsection*{Effects of secondary ecological processes} 

A natural question is whether the marginal nature of Gaussian
competition can be brought on by secondary ecological effects. We
have checked that adding a small immigration rate does not produce
lumpy distributions. Adding noise or an extinction threshold
(i.e.~species are removed when their populations fall below a
threshold) result in a limit to similarity between species, but does
not lead to clustering \citep{Pigolotti2007}. This also happens in non
marginal cases with $p<2$, where the minimum distance between species
is unrelated to the competition range $\sigma$.

\begin{figure}
  \includegraphics[width=8cm]{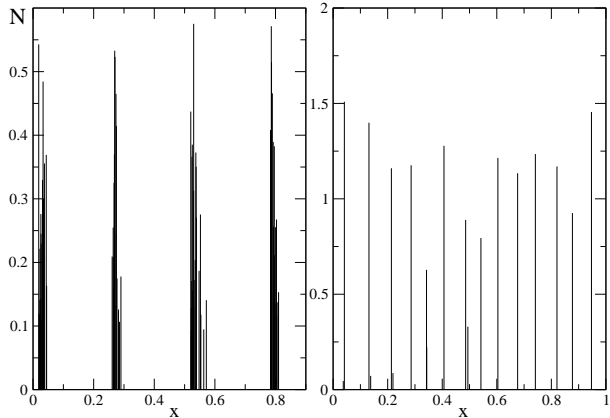} \caption{Final
    populations after $500\,000$ generations with speciation and
    extinction. Species whose population goes below $0.1$ are removed
    from the system. Every $100$ generations new species are
    introduced close to an existing one. The parent species is chosen
    with a probability proportional to its population; the distance of
    the new species to its parent is drawn from a Gaussian
    distribution of zero mean and spread $\sigma_p=0.02$. The new
    species $j$ is introduced with a population uniformly drawn from
    the interval $N\in [2,3]$.  (Left panel) Gaussian kernel ($p=2$)
    and (Right panel) exponential kernel ($p=1$). Simulations are
    performed with fully periodic boundary conditions. $K=10$ and
    $\sigma=0.3$.}
  \label{fig:immigration}
\end{figure}

Effect of species extinction, invasion, and speciation was simulated
by eliminating species below a given population threshold, and
introducing invading species at a fixed rate.  If they are introduced
at random locations in niche space no patterns are observed. If
invading species are introduced close to existing ones, modeling
evolutionary change and speciation \citep{Lawson}, the system ends
with a lumped species distribution, even for $p=2$
(Fig.~\ref{fig:immigration}). However, the same mechanism has no
effect if an exponential competition kernel ($p=1$) is chosen. The
interpretation is that evolutionary effects favor the formation of
lumpy species distributions, but only when the competition kernel is
close to the Gaussian limiting case.

\subsection*{Effects due to truncation}

\begin{figure}
  \includegraphics[width=8cm]{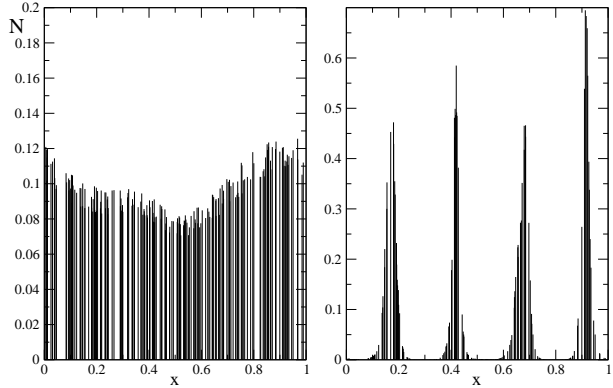}
  \caption{Populations of $200$ species after $1000$ generations with
    (left panel) Gaussian competition kernel with fully periodic
    boundary conditions, and (right panel) with truncated Gaussian
    competition kernel (see text). $K=10$ and $\sigma=0.3$. }
\label{fig:numerical}
\end{figure}

The most obvious numerical simplification is to only partially
implement the periodic boundary conditions, by omitting the kernel
wrap around the niche interval, that is, using $G(y)$, with $y$ being
the minimum of the two possible distances among species $i$ and $j$
($|x_i-x_j|$ and $1-|x_i-x_j|$), instead of the periodic kernel
$G_p(y)$. The resulting effective kernel is Gaussian but truncated at
$|y|=1/2$, making it no longer positive definite. Although the shapes
of $G(y)$ and $G_p(y)$ are still very similar for the parameters used
here ($\sigma=0.3$), the change immediately leads to lumped species
distributions (Fig.~\ref{fig:numerical}). In contrast, for $p=1$ (or
any other values of $p<2$ which we have checked), changing $G_p(y)$ by
$G(y)$ has no noticeable effect.  Qualitatively, the dynamics for
truncated Gaussian kernels resembles the outcome when the exponent of
the competition kernel is perturbed just slightly.  E.g.~using $p=2.1$
instead of $p=2$ also leads to lumped species distributions, even when
fully periodic boundary conditions are implemented (not shown). 

While the mathematical analysis of ``evolutionary diffusion'' is
rather complicated and results can be obtained only in the framework
of simple models \citep{Lawson}, the effect of truncation on the
different kernels may be analyzed in more depth and is discussed in
the next subsection.

\subsection*{Wavelength of Gaussian and exponential
  instabilities}

We have numerically shown in the previous section that truncation
leads to radically different results for the Gaussian and exponential
kernel. This may be surprising when realizing that neither the
truncated Gaussian nor the truncated exponential kernel are positive
defined.  The explanation of the different result comes from the
wavelength of the modes $k$ for which the Fourier transform of the
kernel $\tilde G(k)$ takes negative values, as it determines the
distance between lumps \citep{Pigolotti2007,Fort2009}.

To find these wavelengths, we start from the Gaussian case and show
what is the effect of truncating the kernel at a distance $a$ (which
we take now to be the niche-space size, previously scaled to be 1) :
\begin{equation}
 \tilde G_{p=2}(k)= \int_{-a}^a \exp(-x^2/\sigma^2-ikx)\ dx=\frac{\sqrt{\pi} \sigma} {2}
  e^{-\frac{k^2\sigma^2}{4}} \left [\mathrm{erf}\left(\frac{a}{\sigma} + 
\frac{ik\sigma}{2} \right)+c.c.\right],
\end{equation}
where $\mathrm{erf}(x)=(2/\sqrt{\pi})\int_0^x \exp{-t^2}dt$ is the
error function, and $c.c.$ stands for complex conjugate.  We impose
$a/\sigma\gg1$, so we can get a simpler expression by
expanding the error function, yielding
\begin{equation}
  \tilde G_{p=2}(k)\approx\frac{\sqrt{\pi}\sigma}{2}\left[e^{-\frac{k^2\sigma^2}{4}}
    -2e^{-\frac{a^2}{\sigma^2}}\left(\frac{\sigma\cos(ka)}{\pi a}-
      \frac{2\sin(ka)}{\pi k\sigma}\right)\right]
\end{equation}
From the previous expression we see that the wavenumbers that give
negative modes are order $k\approx 2a/\sigma^2$.

We now check the same effect in the exponential case:
\begin{equation}
\tilde G_{p=1}(k)=\int_{-a}^a \exp(-|x|/\sigma-ikx)\ dx=\frac{2\sigma\left(1-e^{-\frac{a}{\sigma}}
\cos(ak)+ e^{-\frac{a}{\sigma}}k\sigma \sin(ak)\right)}{1+k^2\sigma^2} 
\end{equation}
In this case the instability occurs for large wavenumbers,
proportional to $\exp(a/\sigma)$. We checked numerically that kernels
with $p<2$ behave like the $p=1$ case, with the wavenumber of the
first negative mode growing exponentially with the size of the niche
space.

Summarizing, the kernels we considered develop an instability due to
the truncation, but at very different frequencies: very high for the
exponential kernel (exponential in the ratio between the system size
and the kernel range) compared with the Gaussian (proportional to the
same ratio).  The consequence is that, already when the size of the
niche space is large but not extremely large compared with the
competition distance (like the cases considered here, $\sigma/a=0.3$),
the unstable mode in the exponential case has a wavelength being much
smaller than the interspecies distance, and is thus unable to generate
patterns.

\section*{Discussion}
The model (\ref{eq:LV})-(\ref{eq:alpha}) provides a very abstract
representation of competition. Both empirical observations and
theoretical approaches, based on explicit consideration of the coupled
consumer-resource dynamics, lead to competition coefficients which are
quite different from Gaussian
\citep{Schoener1974,Wilson1975,Ackermann2004}, except in a few
particular cases. Even so, the qualitative outcome of the model does
not depend on the exact shape of the competition kernel, but only on
its positive or non-positive definite character. We have restricted
our considerations primarily to the basic model (\ref{eq:LV}) with
Gaussian interaction kernel and constant carrying capacity since it is
the simplest implementation, allowing to illustrate in a clear setting
the importance of $G$ and the issues caused by the choice of a
marginal competition kernel.

The basic model with competition coefficients obtained from the
overlap of utilization functions, which yields always positive
definite kernels, allows for dense species distributions with no
limits to similarity. This fundamental solution may be changed by
three different effects: 1) effects stemming from the competition
kernel being no longer positive definite lead to lumpy species
distributions. Clusters of species will appear, separated by exclusion
zones in niche space with a spacing proportional to the width of the
competition kernel $\sigma$; 2) second order ecological effects like
noise, species heterogeneity, evolutionary effects, or the
introduction of an extinction threshold lead to a limit to the
similarity with the spacing between species being independent of
$\sigma$; 3) under a non constant carrying capacity, patterns of
unevenly spaced species, lumpy or not, may appear. This lead
\citet{szabo} to conclude that ``the not-very-smooth nature of the
carrying capacity seems to be essential for limiting similarity''.
Notice that also some types of smooth, but non-uniform, carrying
capacities may originate lumped distributions \citep{Garcia2008}

The first case arises when the competition kernel is not positive
definite. The main point of this paper is that, when the competition
kernel is the marginal Gaussian, the model becomes very sensitive to
additional effects that may lead to lumpy species distribution.  As an
example, we demonstrated that a simple representation of evolutionary
diffusion \citep{Lawson} may lead to patterns in the Gaussian
case. This effect is similar to that of evolutionary dynamics under
assortative mating which shown to lead to lumpy species distributions
\citep{Doebeli2007}. It is worth mentioning that, in the context of
evolutionary dynamics, the fact that the presence of clusters depends
of the choice of the competition kernel has been recently demonstrated
\citep{Leimar2008}. Patterns can also result from numerical
approximations, such as truncating the tails of a Gaussian competition
kernel. This effect is probably the underlying mechanism behind
species clustering observed in recent numerical work
\citep{Scheffer2006}, which was used to explain observed lumpy
distributions \citep{May2007}. Additional ecological effects have been
identified however \citep{Scheffer2006} which make the species groups a
robust phenomenon (See \citet{Garcia2008} for analytical solutions of
this type). In any case, these spurious effects can be avoided by
paying attention to numerical details or by using a competition kernel
which is not marginal, e.g.~one with $p=1.5$, which in practice is
almost indistinguishable from the Gaussian one.  It is worth
mentioning that analytical (i.e.~not numerical) results are not
affected by the marginal nature of the Gaussian kernel, both in
relation to limiting similarity \citep{May1972}, coevolution
\citep{Case1981} or criteria for sympatric speciation
\citep{Doebeli2000}. The marginal nature of Gaussian competition
kernel may however affect numerical work on food web evolution and
assembly \citep{Doebeli2000,Loeuille2005,Lewis2007}.

Since a non-positive definite competition kernel leads to lumpy
species distributions a natural question is whether a non-positive
definite kernel can result from hypothesis on ecological
interactions. This case is often neglected in the literature, since
assuming Eq. (\ref{eq:interaction}) automatically leads to a positive
definite competition kernel \citep{Roughgarden1979}. However, as
emphasized in \citet{Meszena2006} and references therein, under quite
general assumptions one should introduce two different
utilization-like functions: a \emph{sensitivity} function $S_i(x)$,
describing the effect of the resource at $x$ on the growth of species
$i$, and an \emph{impact} function $D_i(x)$, describing the depletion
of resources produced by $i$. Then, the competition coefficients
depend on the overlap of these two quantities $\int S_i(x) D_j(x) dx$,
and reduce to (\ref{eq:interaction}) only if the sensitivity and
impact functions are proportional, with the constant of
proportionality being the ecological efficiency. When the ecological
efficiency is a function of $x$, and the sensitivity and impact
functions are no longer proportional, the competition kernel may cease
to be positive definite \citep{Garcia2008}.

The third mechanism is that of a non-constant carrying capacity
$K(x)$, which has been explored by \citet{szabo}. They found that some
choices of carrying capacity leads to an irregular species
lumping. The effect of non-constant carrying capacity in conjunction
with both positive and non-positive definite competition kernels was
explored by \citet{Garcia2008}. The emerging picture is that the two
mechanisms are independent. The cases in which a non-constant carrying
capacity leads to uniform species distributions can also be
destabilized by a non-positive definite kernel. This means that the
mechanism explored here is not a particularity of constant carrying
capacity but is present also in more general settings.  It has been
suggested that the effect of cutting the kernel is similar to that of
having a finite niche space, i.e. of a box-like carrying capacity
\citep{Fort2009}. However, this analogy does not hold to a rigorous
mathematical analysis, which has demonstrated that lumps may appear
when the carrying capacity is non-constant, but that this effect is
not mathematically equivalent to a cut kernel
\citep{Garcia2008}. Moreover, recent results \citep{Baptestini2009}
show how the effect of closed boundaries depend on the details of the
boundary shape, so that the box-like case is not representative of a
general finite niche space.

Having outlined the reasons that may cause the three different
outcomes, the question arises if it is possible to infer whether one
effect or the other is at play from the result of a numerical
integration of the competition model. It can be difficult to
distinguish between a uniform distribution of discrete species and a
lumpy one when lumps are very narrow and close. Here, the fact that in
the lumpy distribution the spacing of the lumps is proportional to the
width of the competition kernel $\sigma$ can be used. If changing
$\sigma$ results in a change in the distance between species
proportional to $\sigma$, the effect is due to a non-positive definite
competition kernel and vice versa. In the case where the effect is due
to the carrying capacity being non-constant the spacing of species is
usually more irregular \citep{szabo}.

To summarize: in line with previous works we have found that the case
of continuous coexistence (no limits to similarity) may be limited by
a variety of effects, specially for the Gaussian kernel which has a
marginal character.  We have underlined that there are different ways
to limit similarity, some leading to lumpy species distributions and
others not. We hope that this article will increase the awareness in
the theoretical ecological community of the potential subtleties
associated with the use of the Gaussian competition kernel. Even
though this functional form appears to be natural, in particular for
analytical work, it may not be the most prudent choice for numerical
exploration of the niche model.


%
%

\begin{acknowledgements}
C.L. and E.H-G. acknowledge support from project FISICOS
(FIS2007-60327) of MEC and FEDER and NEST-Complexity project
PATRES (043268). K.H.A was supported by the Danish research
council, grant no. 272-07-0485
\end{acknowledgements}


\bibliographystyle{plainnat}

\end{document}